\documentclass[aps,pre,reprint,10pt,superscriptaddress,showpacs]{revtex4-1}
\usepackage{amsfonts} 
\usepackage{amsmath}
\usepackage{amssymb}
\usepackage{graphicx}
\usepackage{subfigure}
\usepackage{color}
\usepackage{caption}
\usepackage{makeidx}
\usepackage{multirow}
\usepackage{ulem}
\newcommand*\xbar[1]{%
  \hbox{%
    \vbox{%
      \hrule height 0.5pt 
      \kern0.5ex
      \hbox{%
        \kern-0.1em
        \ensuremath{#1}%
        \kern-0.1em
      }%
    }%
  }%
} 
\begin{document}
\title{Discrete  model for cloud computing: Analysis of  data  security and data loss}
\author{Animesh Roy}
\email{aroyiitd@gmail.com}
\author{A. P. Misra}
\email{apmisra@visva-bharati.ac.in; apmisra@gmail.com}
\affiliation{Department of Mathematics, Siksha Bhavana, Visva-Bharati University, Santiniketan-731 235,  India}
\author{Santo Banerjee}
\email{santoban@gmail.com}
 \affiliation{Institute for Mathematical Research, Universiti Putra Malaysia, Selangor, Malaysia}
\begin{abstract}
Cloud computing is recognized as one of the most promising solutions to information technology, e.g., for storing and sharing data in the web service which is sustained by a company or third party  instead of storing data  in a hard drive or other devices. It is essentially a physical storage system which provides large storage of data and faster computing to users over the Internet.  In this cloud system, the third party allows to preserve data  of clients or users   only for business purpose and also for a limited period of time. The users are used to share data confidentially among themselves and to store data virtually   to save the cost of physical devices as well as the time.    In this paper, we propose  a discrete dynamical system for  cloud computing   and data management of the storage service between a third party and users.   A framework,  comprised of different techniques and   procedures for distribution of storage and their implementation with users and the third party is given. For illustration purpose, the model is considered for two users and a third party, and  its dynamical properties are briefly analyzed and discussed. It is shown that  the   discrete system   exhibits   periodic, quasiperiodic and chaotic states. The latter  discerns that the cloud computing system with distribution of data and storage  between users and the third party may be secured.  Some issues of data  security are discussed and a random replication scheme is  proposed  to ensure that the data loss can be highly reduced compared to the existing schemes in the literature.               
\end{abstract}
\maketitle
\section{Introduction}
Since it was first commercialized in $2006$ by Amazon's EC2,  cloud-based computing service has become more popular in recent years in information technology (IT) industry because of its performance, accessibility, low cost and many other   advantages  of computing activities including business support \cite{ngenzi2014,sakellari2013,chiregi2017,laili2013}. Cloud computing is an approach in which the users or clients can save their data to an off-site storage system that belongs to a third party and   can access the pool of computing resources as well as the computing power of their own in a network environment.  It also provides a  dynamic, shared and flexible resources from remote data centers to the users, and   supports service request  where the Internet can develop resources to them.
\par
The cloud computing is based on mainly three functional units:
\begin{itemize}
\item[1.]\textbf{Cloud service provider:}  A cloud service provider (CSP)  is a company that offers  network services, infrastructure, or business applications in the cloud. It has a significant storage capacity to preserve the user's data and high computation power, as well as provides the security of stored application.   The large benefit of using a CSP is due to its efficiency, accessibility and  low cost of computing activities.  Furthermore, the individuals and companies, instead of building their own infrastructure to support internal services and applications,  can also purchase the service from the CSP  which provides the same to many customers from a shared infrastructure. There are, however, three main services in the cloud by CSPs,   which can be discussed as follows:\\
\textit{Software as a service} (SaaS): It is the top priority model in which a service  provider (third-party) hosts the software or  application on the cloud infrastructure and makes them available to users over the Internet.  However, the users can not manage or control the underlying cloud infrastructure, network, servers, operating systems or even individual application capabilities except some limited user-specific application configuration settings.  It also helps the users  save cost by renting the application instead of    licensing of the
traditional packages from cloud service.\\
\textit{Platform as a service} (PaaS):  A middle or second category of the cloud service model   that provides a platform or software environment allowing users to design, develop, run, and manage their applications without worrying about  the complexity of building and maintaining the cloud infrastructure.\\
\textit{Infrastructure as a service} (IaaS): It is an instant computing infrastructure,  provisioned and managed over the Internet.  It helps  users avoid the expense and complexity of buying and managing their  own physical servers and other data-center infrastructure. However, the users can purchase, install, configure and manage their own software-operating systems, middle-ware and applications.  
\item[2.]\textbf{Owner/ Third party:} It is   an organization  which stores large data files  in the cloud and relies on it for data maintenance and computation.  
\item[3.]\textbf{User/Client:} It is usually registered with the owner and it uses the data of owner stored in the cloud. The user can be an owner itself as well.  
\end{itemize}
\par 
Data security  is one of the main concerns of cloud computing system as it  encompasses many technologies such as network, database, operating system, memory management etc. So, the service providers must ensure that the   users preserve  their data (i.e., no data loss and data theft) in  storage section without any serious risk. In this context, various security issues and challenges have been addressed in the literature (see, e.g., Refs. \onlinecite{mohamed2013,sood2012,yu,an}). Some architectural security issues are also there which are changing due to various architectural designs over cloud computing. Users primarily want  two kinds of securities:
\begin{itemize}
\item High-end security which provides automated maintenance and support, low cost for all Internet-capable user devices, and individual security settings regardless of operating system and user device.
\item Data security in which all the data is encrypted and sent via the secure system, i.e., no one can intercept  data and security is everywhere even when one uses other devices. 
\end{itemize} 
On the other hand,  the trust issues and challenges (such as \textit{trust in efficiency} and \textit{trust in belief}) are also matters of concern in the cloud computing.    Note that  trust is a significant indicator for service selection and recommendation, and it is still an emerging topic in the cloud computing.  For a comprehensive and systematic review of trust evaluation, readers are suggested to Ref. \onlinecite{chiregi2017}.  Though we are not considering the theory of trust evaluation, however, we will be proposing a cloud computing model from a different point of view which is, in fact, chaotic in nature \cite{banerjee2011,dijk,tobin2017}. This model will be useful for users to share or store data in the cloud through chaotic processes more securely. We will also analyze and discuss various security issues and challenges,  and propose a  scheme for the analysis of data loss, and to show that the probability of data loss can be negligibly small.  
\par
Thus, a dynamical model can be constructed for cloud computing with SaaS    for distribution of data, their storage and security.  The main issue is that one can store and share data or files  in the chaotic regime  without any risk. In this context, several theoretical attempts  have been made from different points of view \cite{ngenzi2014,sakellari2013,chiregi2017}, however, as of now no comprehensive discrete or continuous dynamical model is available in the literature.   Since for a confidential data storage, cloud computing model can be owner independent,   the user section requires a dynamical model which exhibit chaos.  In the proposed  algorithm, to be discussed shortly,  we will see that how at different iterations or discrete times a file is stored randomly for each and every user and securely in the chaotic regime so that  no one can access the files of the others. In the proposed model we also use some scaling parameters for the flexible storage and the independent users, and   fix them  in such a way that the model exhibits chaos for  a system of  two or more users and a single owner.
\par   
 In this work, we propose a discrete dynamical model where a cloud storage operating system provides a set of network-based interfaces i.e., file management and storage which are accessible from various devices of the users. Here, a user can access, save and transfer their files from anywhere (using web, Internet etc.) in a chaotic regime securely and with low cost  and less time. The model exhibits periodic, quasiperiodic and  chaotic states  for certain ranges of parameter values.   Though there are different considerations for cloud computing, we will consider only on a private multi-user single-owner model and it can also be used for multi-owner model as well.     In this process the total data communication system is  shown to be  fully secured, i.e., the main problem of security issues may be reduced. In our model we construct a discrete system in which an owner or third party buys a storage capacity from the cloud storage company and distributes  to the users in such a way that the storage capacity  depends on the authentication of the users.  In this situation the owner can provide a well distribution of the storage as per demand of the users in a chaotic regime. This service is  provided by the owner only, no user and also no cloud storage company can  access without the permission of the owner. We also propose a random replication scheme to ensure that the probability of data loss can be highly reduced compared to the existing scheme in the literature \cite{cidon2013}.   

\section{Proposed model for cloud computing } \label{sec-model}
Mathematical modeling and analysis of cloud computing may  be important  due to understanding of the inter-dependencies of demand of the users and capacity of a owner involved.  
 A cloud storage system requires a  distributed file system  that allows many users to have access to data and supports operations (e.g., create, delete, modify, read and write) on that data, and different file system can have different scale size. In our model, we choose  $n$ users, namely $U_1,~U_2,...,U_n$ and one  owner. We assume that  $v_m$ is the maximum storage capacity of the owner. Also,   let  $x^{(l)}_i$ be the $l$-th demand of the user $U_i$ with a scaling parameter $\xi_i$, and  $v^{(l)}_{c}$ the storage capacity of the owner with    the   scaling parameter $\alpha$ at the $l$-th state or time such that the maximum capacity is $v_m$.  Though the variables  $x^{(l)}_i$ are independent, however,  each of them   depends on the storage capacity $v^{(l)}_{c}$.  Here, the owner with storage capacity $v^{(l)}_{c}$ is flexible with respect to  the user's demand $x^{(l)}_i$. 
  In the cloud computing system, an owner  buys a maximum storage capacity $v_{m}$ from a cloud service provider  as per  users'   flexible demands, and  requires a   dynamical model for distribution and share of data through a random process in which a chaotic regime exists where the users' files  may be encrypted by the public key RSA algorithm and their passwords may be secured by the Hash function.     Thus,  we can construct a discrete model which will describe  sharing of data, distribution of storage from owner's section to every user $U_i$ as per their demand $x^{(l)}_i$ at any $l$-th state or $l$-th discrete time.  
\par 
In the  process of cloud computing a storage device is connected with  the owner and initially a storage size is distributed  with the capacity variable $v^{(l)}_{c}$ to the users as per   their demands.  Since $\alpha$ is the scaling parameter for the capacity variable $v^{(l)}_{c}$,  the maximum storage capacity $v_m$  of the owner is such that $\alpha v^{(l)}_{c} \leq  v_m$, where    $0<\alpha\lesssim1$.  Also, the sum of the  storage distributions to the $n$ users at any stage $l$ is always less than the maximum capacity of the owner $v_{m}$ i.e.,   
\begin{equation}
0 < \sum_{i=1}^{n} (-1)^i  \xi_i x^{(l)}_i \leq \alpha v^{(l)}_c \leq v_{m}, \label{eq-param-rest}
\end{equation}
where  $\sum_{i=1}^{n}(-1)^i \xi_i \leq 1$ and $\xi_i\geq0~\forall~ i$.  The minus sign in Eq. \eqref{eq-param-rest} is introduced to measure the  demand or storage of an user relative to the other.  
Next, we  measure the changes in the proposed system at any state $l$.  In the distribution of storage from an owner to the users,  the   capacity variable at $(l+1)$-th state, $v^{(l+1)}_{c}$ must be equal to the  capacity in the  $l$-th state (with the unit scale  $\alpha$) minus the  storage already allocated as per the demands of the users (with the unit scale $\xi_i$). Thus, the capacity in the owner's section in the $(l+1)$-th state is given by
 \begin{equation}
 v_c^{(l+1)}=\alpha v^{(l)}_{c} -\sum_{i=1}^{n}(-1)^i\xi_i x^{(l)}_i, \label{eq-vc}   
\end{equation} 
where $l=0,1,2,...$ 
On the other hand, the demand of an user also varies, in general, from $l$-th state to $(l+1)$-th state. Since at any state $l$, $v^{(l)}_{c}$ denotes the percentage capacity of $v_m$,  the demand of an user $U_i$ at $(l+1)$-th state depends on the storage availed by the other users $U_j$  $(j\neq i)$ and the storage of the user $U_i$ in the $l$-th state.
Thus, we have for $l=0,1,2,...$ and $i=1,2,...,n$
\begin{equation}
 x_i^{(l+1)}=(-1)^i\xi_i x_i^{(l)} v_{c}^{(l)} -\sum_{j=1;j\neq i}^{n}(-1)^j\xi_j x_j^{(l)}. \label{eq-xi}.
\end{equation}
 At $l=0$, we have certain initial values of the capacity and the demand variables, i.e.,  $v_{c}^{(0)}$ and $x_{i}^{(0)}$. We assume that initially   with the   capacity $v_{c}^{(0)}$, the company or owner  uniformly distributes to all the users the amount of storage $x^{(0)}$. As an illustration, we analyze our model for one owner and two users.  Thus,  Eqs. \eqref{eq-vc} and \eqref{eq-xi}  reduce to
\begin{equation}
\begin{split}
   & v_{c}^{(l+1)}= \alpha v_{c}^{(l)}+\xi_1 x_1^{(l)}-\xi_2 x_2^{(l)}, \\
   &x_1^{(l+1)}=-\xi_1 x_1^{(l)} v_{c}^{(l)}-\xi_{2} x_2^{(l)},\\
 &x_2^{(l+1)}=\xi_1 x_1^{(l)} + \xi_2 x_2^{(l)} v_{c}^{(l)}, \label{eq-model}
 \end{split}
 \end{equation}
where $l=0,1,2,...$; $x_i,~\xi_i\geq0$ for $i=1,~2$ and $0<\alpha\lesssim1$. 
\par
\begin{figure*}
    \centering
    \subfigure[]
    {
        \includegraphics[width=4in,height=2.5in]{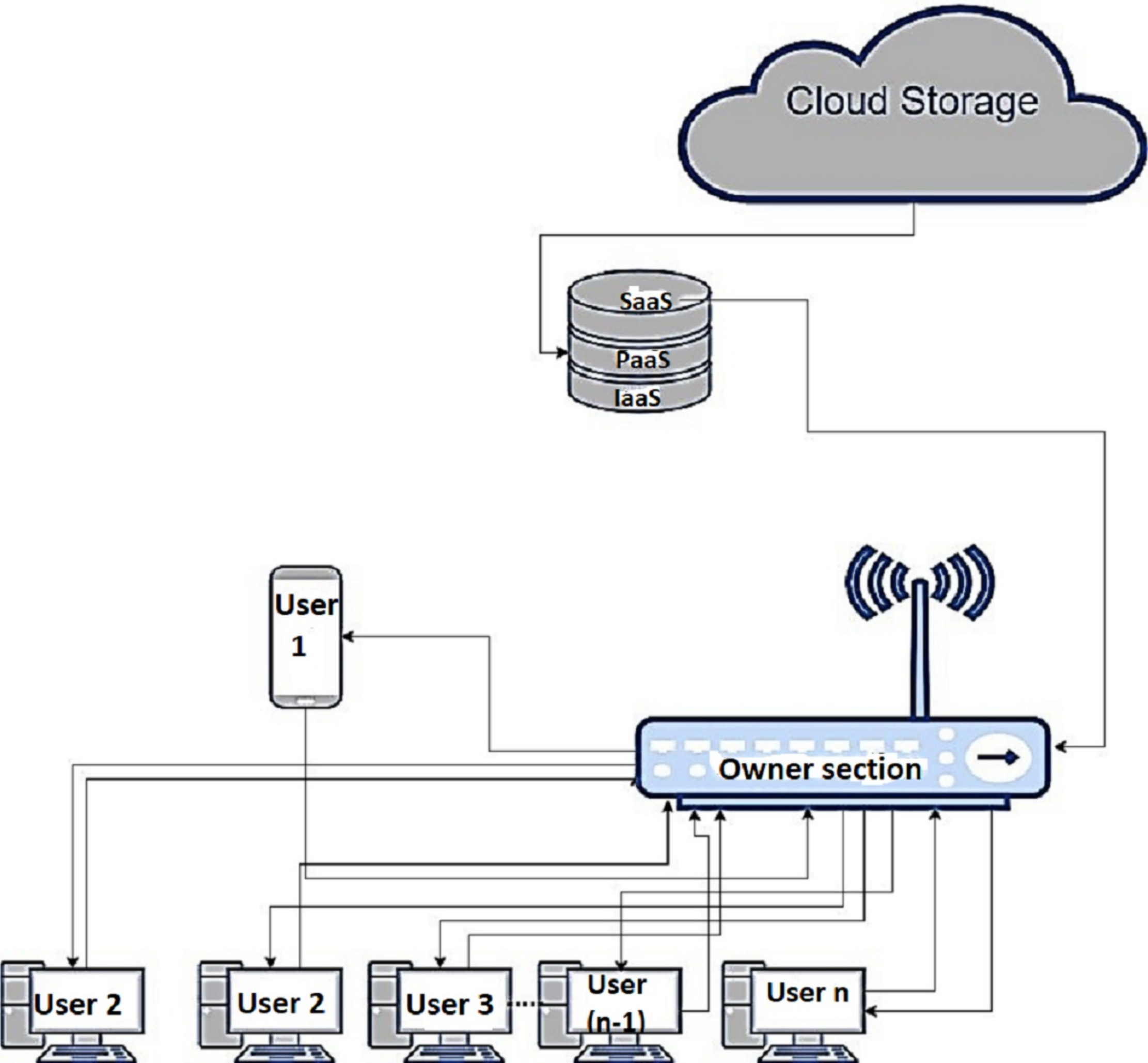}
    }
    \\
    \subfigure[]
    {
        \includegraphics[width=4in,height=2.5in]{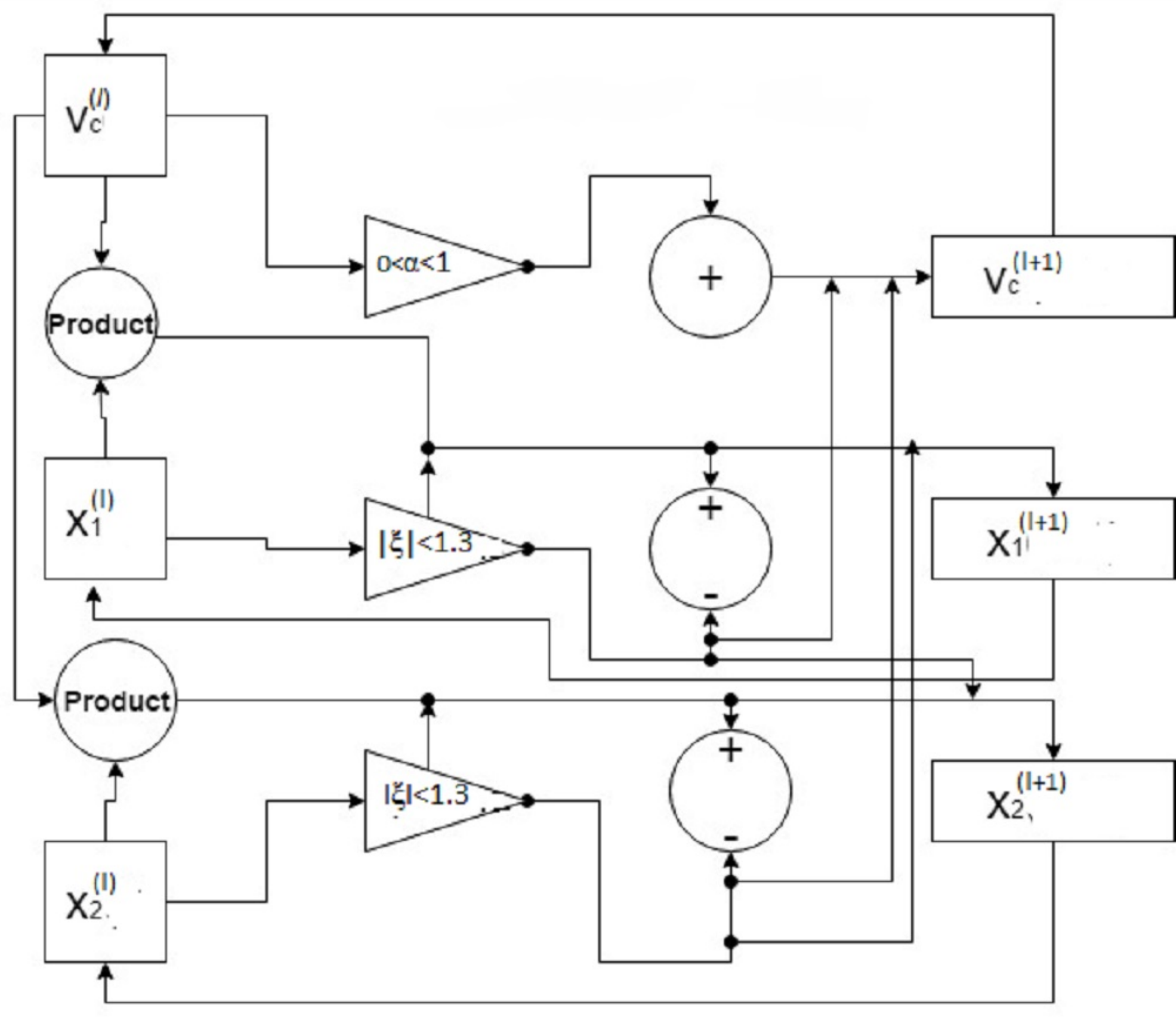}
    }
    \qquad
    \caption
    { A schematic diagram [panel (a)] and a description of the model [panel (b)] for the process of cloud computing. 
    }
    \label{fig:schm-diagram}
\end{figure*}

  A schematic diagram for a cloud computing model [panel (a)] and  its  description   [panel (b)] as in Sec. \ref{sec-model} are shown in Fig. \ref{fig:schm-diagram}.  The   model  can be implemented with the   following steps.
\begin{itemize}
\item[1.] Assume the initial storage capacity of the    owner as  $v_c^{(0)}$ at  $l=0$  and distribute it uniformly to the users  with demands ~$x_1^{(0)},~x_2^{(0)},...,x_n^{(0)}$.
\item[2.] Consider  the proposed dynamical model with the initial conditions $x_i^{(0)}$, $v^{(0)}_c$ for $ i=1,2,...,n$.
\item[3.] After a certain stage or discrete time $l$,  distribute the storage from the owner's section    to users as per   the flexible storage variable   $x_i^{(l)}$ and the capacity variable $v_{c}^{(l)}$. At the next iteration, i.e., in the $(l+1)$-th state,  the corresponding variables are different from $x_i^{(l)}$ and $v_{c}^{(l)}$,  and they are $x_i^{(l+1)}$ and  $v_{c}^{(l+1)}$  respectively.
\item[4.]  The owner notes  the distributions of  storage to the users  at each state $l$ and makes a secure communication between them.  Any user does not get any kind of information of the other user. 
\end{itemize}
\section{Dynamical properties of the proposed model } In this section, we study the dynamical properties of the  proposed model \eqref{eq-model} for a single owner or company and two users.  First, we perform the linear stability analysis of the model about the fixed points, and second, we show that the chaos may be established by the analysis of Lyapunov exponent spectra and the bifurcation diagram.  
\subsection{Stability analysis of the proposed model} \label{sec-sub-stabi-analy} 
\begin{figure}[htbp]
    \centering
     \subfigure[]{
            \includegraphics[width=2.8in,height=2.0in]{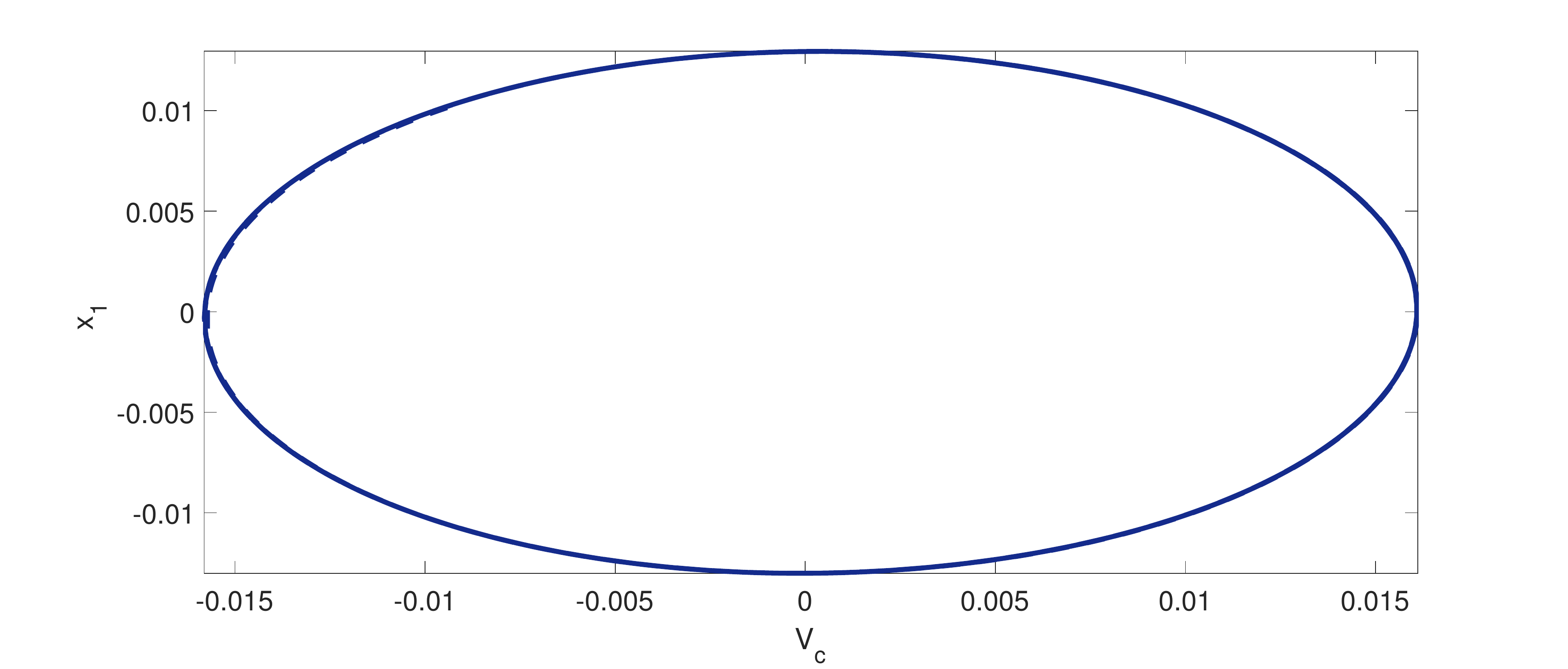}
            }
        
    \subfigure[]{ 
        \includegraphics[width=2.8in,height=2.0in]{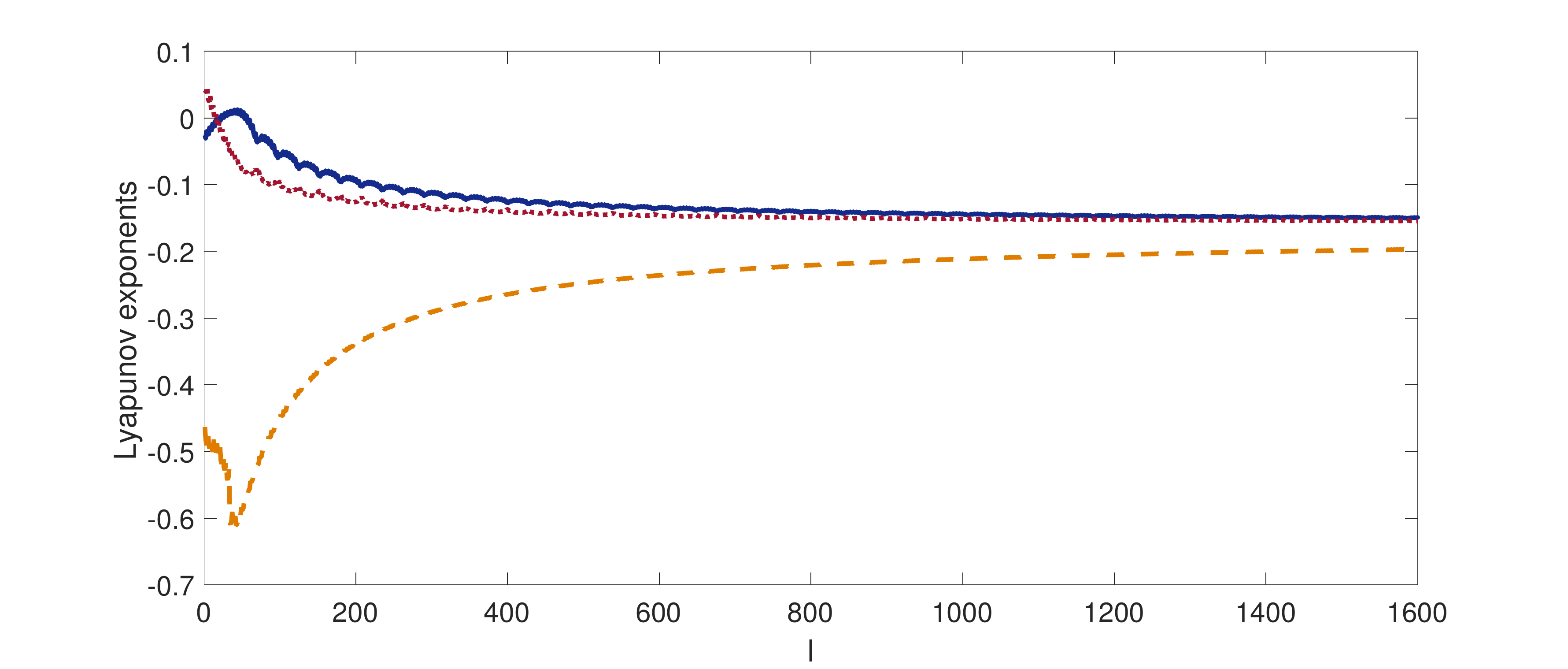}
        }
        
        \caption{ The periodic limit cycle [panel (a)] and the corresponding Lyapunov exponents (all are negative)  [panel (b)] are shown for  $0<\alpha\equiv0.96<\xi_2-\xi_1\equiv0.98<1$.   }
        \label{fig:lperiodic}
        
\end{figure}

\begin{figure}[h!]
    \centering
   \subfigure[]{
        \includegraphics[width=2.8in,height=2.0in]{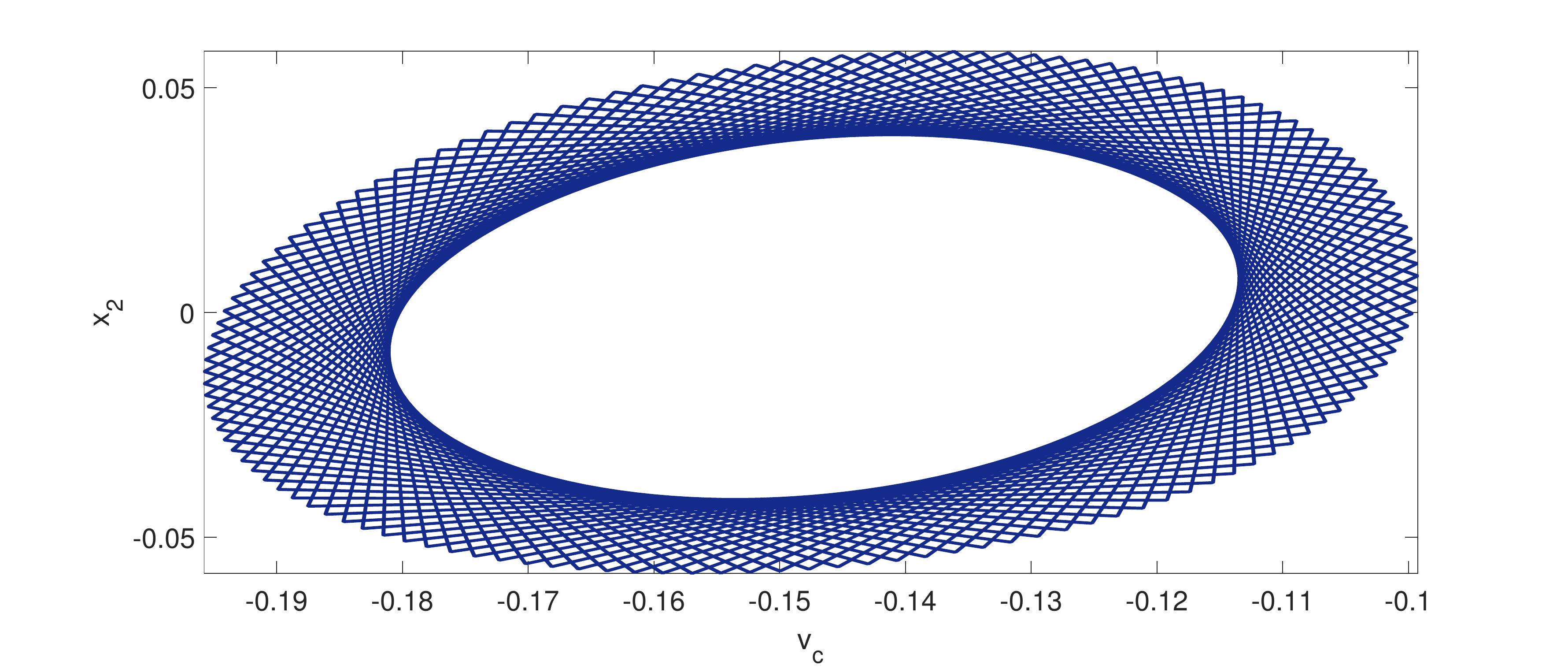}
        }
  
   \subfigure[]{ 
        \includegraphics[width=2.8in,height=2.0in]{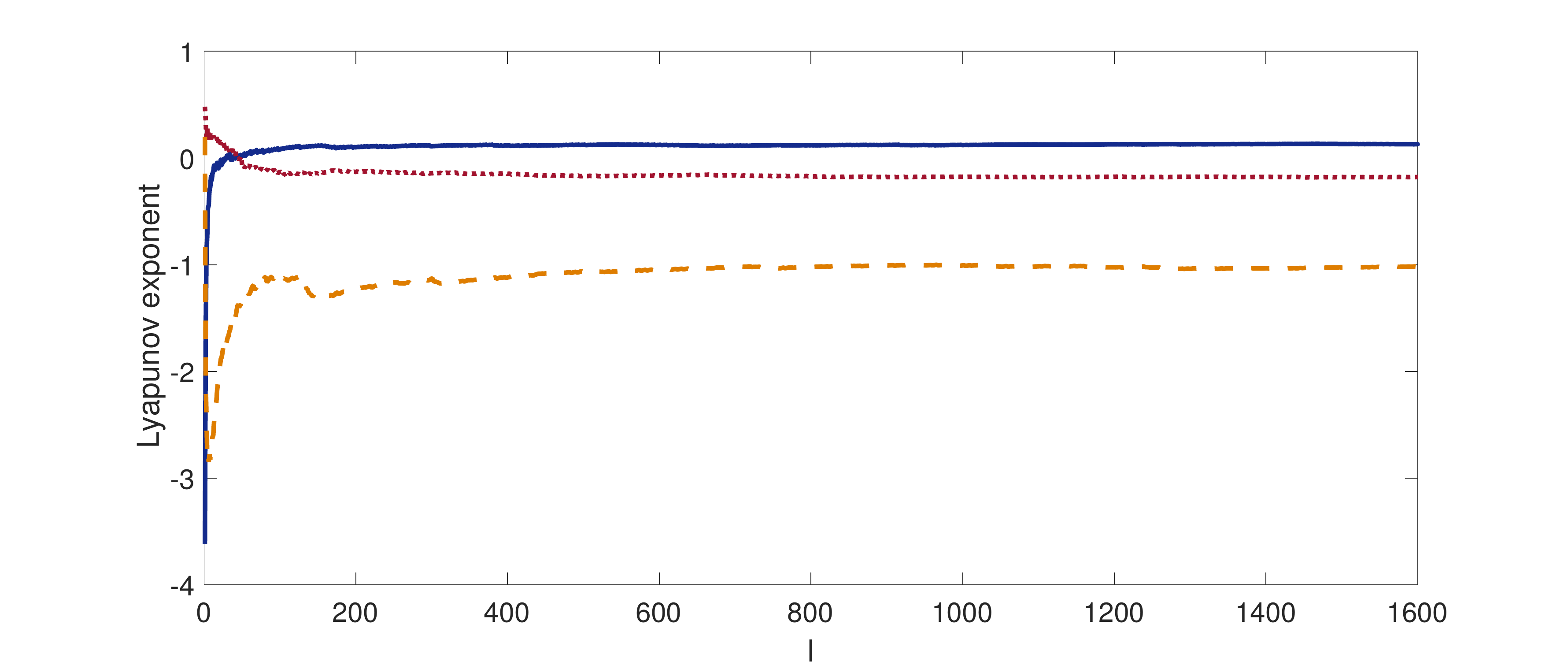}
        }
        \caption{The torus [panel (a)] and the corresponding Lyapunov exponents (one is close to zero and other two are negative) [panel (b)] are  shown  for  $0 <\xi_1-\xi_2\equiv 0.6<\alpha\sim0.9$; $\xi_1=1.4,~\xi_2=0.8$. }
        \label{fig:ltorus}
  
\end{figure}

\begin{figure}[h!]
    \centering
    \subfigure[]
    {
        \includegraphics[width=2.8in,height=2in]{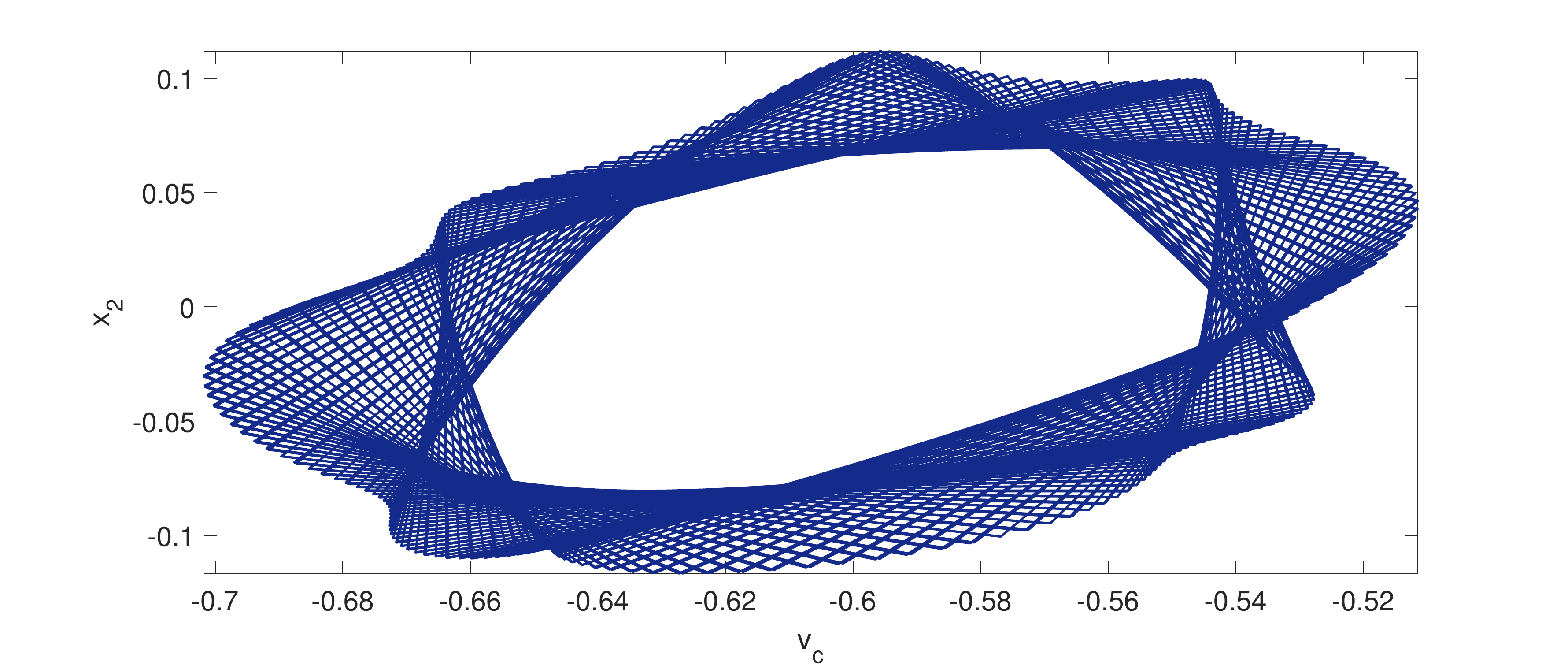}
        }
  
\subfigure[]
    { 
        \includegraphics[width=2.8in,height=2in]{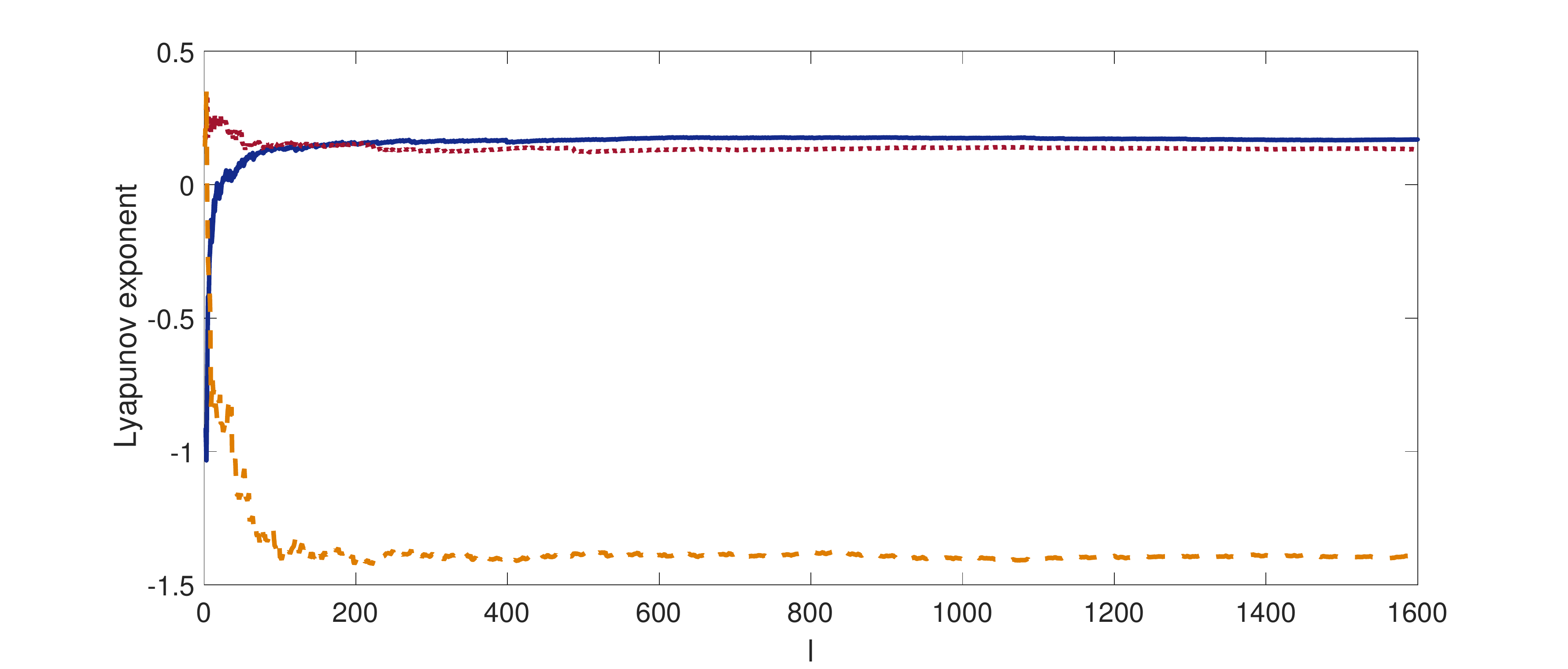}
        }
        \caption{Transition from torus to chaos [Panel (a)]   and the corresponding Lyapunov exponents (two are positive and other one is negative) [Panel (b)] are shown  for  $0<\xi_1-\xi_2\equiv0.05<\alpha\sim0.6$.}
        \label{fig:lchaos}
  
\end{figure}

\begin{figure*}[h!]
    
        \includegraphics[scale=.5]{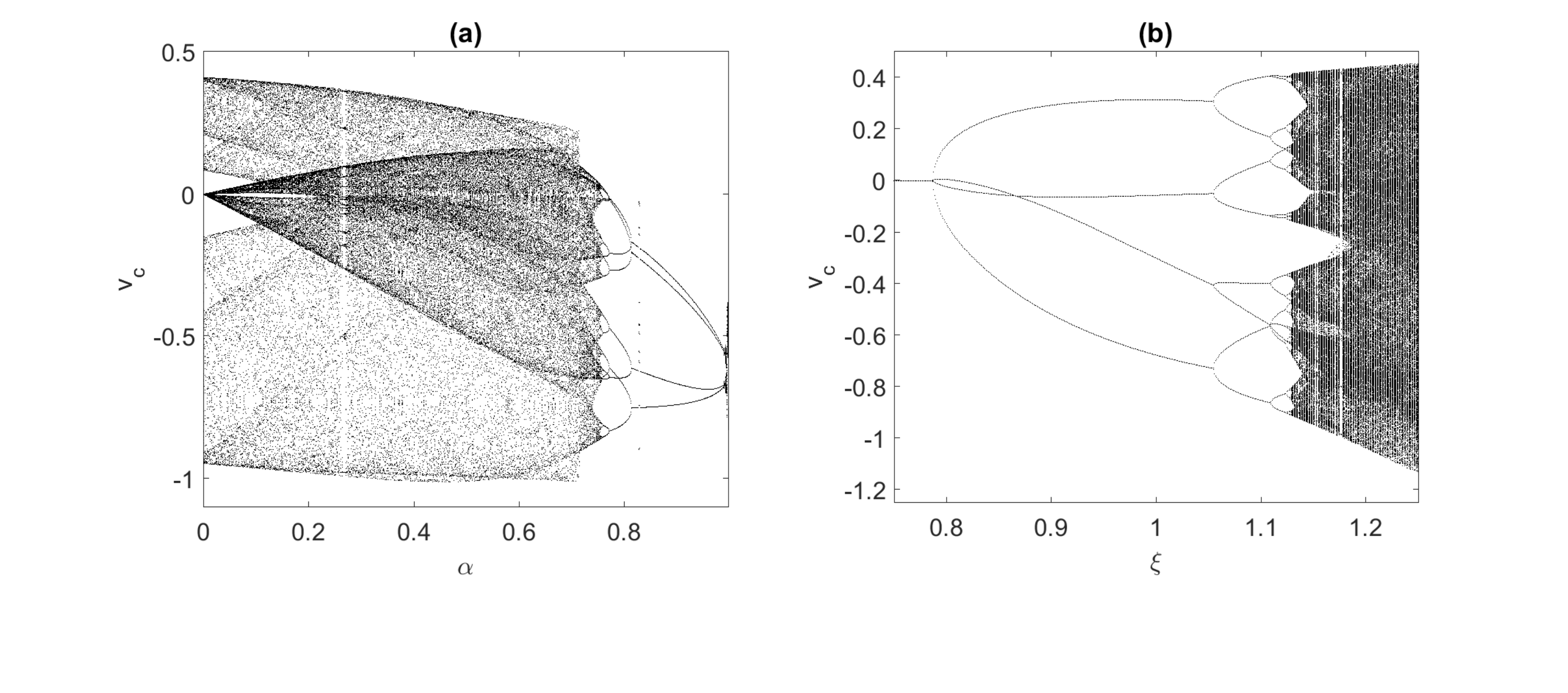}
        \caption{Bifurcation diagram for the capacity variable $v_c$  with respect to the scaling  parameters $\alpha$  and $\xi_1$. Panels (a) and (b) are corresponding to the parameters $(\xi_1=1.28,~\xi_2=1.23)$ and $(\xi_2=1.28,~\alpha=0.5)$  respectively.   }
        \label{fig:bifurcation}
  
\end{figure*}

We study the stability of the fixed points of the iterated map \eqref{eq-model}    with   certain parameter values. 
The fixed points of Eq. \eqref{eq-model}   are obtained as $(0,0,0)$ and $(1,{-\alpha}/{2\xi_1},{\alpha}/{2\xi_2})$, where $\alpha,~\xi_1,~\xi_2\geq0$. Then the eigenvalues of the Jacobian matrix, given by,  
\begin{equation}
  J=\begin{bmatrix}
\alpha & \xi_1 & -\xi_2 \\
0 & 0 &-\xi_2 \\
0 & \xi_1 & 0
\end{bmatrix}
\end{equation}
corresponding to  the fixed point $(0,0,0)$,  are $\lambda_1=\alpha$ and $\lambda_2=\pm\sqrt{-\xi_1\xi_2}$, i.e., $\lambda_1>0$ and $\lambda_2$ is purely imaginary. This means that   the equilibrium is  unstable and there is a possibility of Hopf bifurcation. From the bifurcation diagram (\textit{cf}. Fig. \ref{fig:bifurcation}) we will see  that a critical point indeed exists at  which  the system looses its stability leading to chaos. The latter is established by the analysis of Lyapunov exponent spectra [\textit{cf}. panels (b) of Figs. \ref{fig:lperiodic} to \ref{fig:lchaos}].       
\par
 Next,  we study the stability of the fixed point $(1,{-\alpha}/{2\xi_1},{\alpha}/{2\xi_2})$. To this end, we apply the linear transformations as follows:
\begin{equation}
\begin{split}
 &\widehat{v}_c^{(l)}\longleftarrow \left(v_c^{(l)} - 1\right),~
  \widehat{x}_1^{(l)}\longleftarrow \left(x_1^{(l)} + {\alpha}/{2\xi_1^{(l)}}\right),\\  
 & \widehat{x}_2^{(l)}\longleftarrow \left(x_2^{(l)}-{\alpha}/{2\xi_2^{(l)}}\right).
  \end{split}
  \end{equation} 
Thus, Eq. \eqref{eq-model} reduces to the following linearized form 

\begin{equation}
\begin{split}
 &\widehat{v}_c^{(l+1)} = \alpha\widehat{v}_c^{(l)} +\xi_1 \widehat{x}_1^{(l)}- \xi_2\widehat{x}_2^{(l)},\\
&\widehat{x}_1^{(l+1)}= -\xi_1\left(\widehat{x}_1^{(l)} - \frac{\alpha}{2\xi_1}\widehat{v}_c^{(l)}\right)- \xi_2\widehat{x}_2^{(l)},\\
& \widehat{x}_2^{(l+1)}=\xi_1 \widehat{x}_1^{(l)}  + \xi_2\left(\widehat{x}_2^{(l)}+\frac{\alpha}{2\xi_2}\widehat{v}_c^{(l)}\right). 
\end{split}
\end{equation}
The corresponding Jacobian matrix about the point   $\widehat{O}(0,0,0)$   becomes 
\begin{equation}
J^{*}= \begin{bmatrix}
\alpha & \xi_1 & -\xi_2 \\
{\alpha}/{2} & -\xi_1 & -\xi_2 \\
{\alpha}/{2} & \xi_1 & \xi_2
\end{bmatrix}. \label{jacobian1}
\end{equation} 
The eigenvalues corresponding to the matrix $J^*$ are given by 
 \begin{equation}
  \begin{vmatrix}
\alpha - \lambda & -\xi_1 & -\xi_2 \\
{\alpha}/{2} & \xi_1 - \lambda & -\xi_2 \\
{\alpha}/{2} & -\xi_1 & \xi_2 - \lambda
\end{vmatrix}=0. \label{jacobian2}
\end{equation}
This gives
 \begin{equation}
\lambda^3 +P\lambda^2 + Q\lambda +R =0, \label{eq-lambda}
\end{equation}
where the coefficients are 
\begin{equation}
P = -(\alpha-\xi_1+\xi_2),~Q = \frac{3}{2}\alpha(\xi_2-\xi_1),~R = 2\alpha\xi_1\xi_2.
\end{equation}
  Now, if $P,~Q,~R > 0$  and   $PQ > R$, then by the    Routh's stability criterion  all the roots  of Eq. \eqref{eq-lambda} have negative real parts implying that   the linearized system is stable.   Thus, for the linear stability of the system \eqref{eq-model} about the fixed point $(1,{-\alpha}/{2\xi_1},{\alpha}/{2\xi_2})$, we must have     $0<\alpha<\xi_2-\xi_1\leq1$. Otherwise,   if at least  one of $P$, $Q$ and $R$ is negative (which holds when  $0<\xi_1-\xi_2<\alpha$) and    $\Re \lambda  > 0$, the system is said to be unstable (\textit{cf}. Figs. \ref{fig:lperiodic} to \ref{fig:lchaos}).   Furthermore, in order to have a Hopf bifurcation of the system about the fixed point (for which the equation of $\lambda$ has a pair of purely imaginary roots),  one must have   $PQ=R$, i.e.,   $\alpha = \left[3(\xi_1- \xi_2)^2+4\xi_1\xi_2\right]/3(\xi_1 - \xi_2)$, which gives $|\alpha|>1$. This is not admissible as  $\alpha$ is assumed to vary in $0<\alpha\leq1$. Thus, in this case, there is no possibility   Hopf bifurcation.
\par
{\textit{\textbf   {Calculation of Lyapunov exponents:}}}
In order to calculate the Lyapunov exponents of the discrete dynamical system \eqref{eq-model}, we redefine the variables  as $X^{(l)}=\left(v_{c}^{(l)},~x_1^{(l)},~x_2^{(l)}\right)$ and  the initial condition   as  $X^{(0)}=\left(v_{c}^{(0)},~x_1^{(0)},~x_2^{(0)}\right)$. Then Eq.  \eqref{eq-model} can be rewritten as   
\begin{equation}
X^{(l+1)}= B^{(l)}X^{(l)},
\end{equation}
where\begin{equation}
 B^{(l)}=
\begin{bmatrix}
\alpha &\xi_1  &-\xi_2 \\
0 &-\xi_1v^{(l)}_{c} &-\xi_2 \\
0 &\xi_1 &\xi_2v^{(l)}_{c} \\
\end{bmatrix}.
\end{equation}
Next,  to calculate the Jacobian  $J^{(l)}$ of the above equation for determining the eigenvalues, we differentiate the right-hand side of Eq. \eqref{eq-model} with respect to the variables $X^{(l)}$. Thus, we write
\begin{equation}
  J^{(l+1)}=A^{(l)}J^{(l)}. \label{eq-lyapunov}
\end{equation}
where $A^{(l)}$ is given by
\begin{equation}
A^{(l)}=\begin{bmatrix}
\alpha &\xi_1  &-\xi_2 \\
-\xi_1x^{(l)}_1 &-\xi_1v^{(l)}_{c} &-\xi_2 \\
\xi_2x^{(l)}_2 &\xi_1 &\xi_2v^{(l)}_{c}\\
\end{bmatrix}.
\end{equation}
Furthermore, for $l=0$,  $J^{(0)}$ is the identity matrix and $A^{(0)}$ is evaluated with the initial condition $X^{(0)}$.  So, the solution of the  Eq. \eqref{eq-lyapunov}   is given by 
\begin{equation}\Lambda=\lim_{l\rightarrow \infty}\frac{1}{2l}\log\left(J^{(l)}\left[J^{(l)}\right]^{T}\right),
\end{equation}
 where the Lyapunov exponents are obtained as the eigenvalues $\lambda_i$ of the matrix $\Lambda$ for  $i=1,2,3$. Next, substitution of the   initial condition $X^{(0)}=(0.01,0.01,-0.01)$ and different parameter values of $v_{c}$, $x_1$ and $x_2$   gives   different eigenvalues. 
%

\par
In order to verify the results with the linear stability analysis as above, we numerically solve Eq. \eqref{eq-model} by the fixed point iteration scheme within the ranges of   values of the parameters $\alpha$, $\xi_1$ and $\xi_2$ where the system exhibits stable and unstable fixed points. We consider   the initial condition   as $v_{c}=0.01$, $x_1=0.01$ and $x_2=-0.01$.  The results are displayed in Figs. \ref{fig:lperiodic} to  \ref{fig:bifurcation}. Figure \ref{fig:lperiodic} clearly shows that when the condition $0<\alpha<\xi_2-\xi_1$ is satisfied for the stability of the fixed point $(1,{-\alpha}/{2\xi_1},{\alpha}/{2\xi_2})$, the system indeed exhibits a stable structure (periodic limit cycle) for $\alpha=0.96$ and $\xi_2-\xi_1=0.98$ [see panel (a)]. This means that when   $\alpha \approx 1$, the whole system is running through  the owner's capacity   and the users demands periodically. It happens only when the owner   has almost utilized the capacity and the users are filled with their demands, i.e., no demand after a certain stage or time remains. The periodic structure is justified with the corresponding Lyapunov exponents which are all negative as expected [see  panel (b)]. 

\par
On the other hand, when $0<\xi_1-\xi_2<\alpha<1$, i.e., when the owner has the variable storage (high or low or in between), the periodicity breaks down as predicted from the linear analysis. The corresponding phase-space portraits and Lyapunov exponents are shown in the upper and lower panels of Figs. \ref{fig:ltorus} to \ref{fig:lchaos}. Figure \ref{fig:lchaos} shows that as the value of $\alpha$ is reduced from $\alpha=0.9$    (Fig. \ref{fig:ltorus}) to $\alpha=0.6$    with a different set of values of $\xi_1$ and $\xi_2$, a transition from  torus structure to chaos occurs. The phase-space structures [panels (a)]    are well justified with the    Lyapunov exponents  [panels (b)] as shown in   Figs. \ref{fig:ltorus}   and \ref{fig:lchaos}.  

%
  
~~
\par
\textbf{ \textit{  Bifurcation diagram:} }
Figure \ref{fig:bifurcation} shows the bifurcation diagram   for the capacity variable $v_c$ with respect to the scaling parameters $\alpha$ [left panel (a)] and $\xi$ [right panel (b)]. We find that for some fixed values of $\xi_1$ and $\xi_2$, as the value of $\alpha$ increases in $0<\alpha\lesssim1$, the system approaches from chaotic state to a periodic state with a series of period-halving bifurcations [panel (a)]. This is  a consequence of  Fig. \ref{fig:lperiodic} where the system's periodicity is shown for a value of $\alpha$ close to the unity.  However, an opposite trend occurs when  a fixed value of $\alpha=0.5$ is considered and   one of $\xi_1$ and $\xi_2$ varies [see panel (b)]. In this case, a series of period-doubling  bifurcations leads the system from order to chaos.    Thus, whether the system exhibits periodicity or chaos that depends on the scaling parameters $\alpha,~\xi_1$ or $\xi_2$.
\section{Analysis of data  security and data loss} 
 Data  security \cite{mohamed2013,sood2012,yu,an} and data loss \cite{cidon2013,dataloss} are two main challenging issues in cloud computing systems.  Though, a number of authors have considered these issues in their different works, however, we review the security issues, especially, loss of user identity and password, brute force attack, unauthorized  server, data privacy etc. to justify how our proposed model fits to resolve these security issues. Furthermore, we propose a random replication scheme which  ensures that the probability of data loss is highly reducible compared to the existing scheme in the literature \cite{cidon2013}.  
 There are  two main issues in the cloud computing system, these are
\begin{itemize}
\item A model of organization has to be built up with a  company (owner) and users, and distribution of   storage must be in a proper way and secured.
\item Once the storage is distributed properly, one must be ensured that the  data loss or data leakage is minimum.  
\end{itemize}   
The first issue is resolved by    proposing a chaotic dynamical model which we have already done in Sec. \ref{sec-model}. However,  we analyze and discuss about the second issue, i.e., the data security and    data loss in Secs. \ref{sec-sub-data-security} and \ref{sec-sub-data-loss}.
 \subsection{Analysis of data  security} \label{sec-sub-data-security}
 We  start by considering an example  that a financial company  provides a data storage system to
each of its users  as per their demands in a certain interval of discrete time. The details are demonstrated in Table   \ref{table}. The  model \eqref{eq-model} gives an information about how much data an user can store  in the time interval. 
 We assume that the  company   or owner uses the model for data storage as SaaS. In the following, we define different terms that are related to computation of storage in the user section in  a given   interval of time or at a particular state.
\begin{itemize}
\item \textit{Time interval for demand: } In the model \eqref{eq-model}, we consider the iteration $l$ as some particular time.
\item \textit{Data in use: } In a cloud computing system, the full storage in the owner's section  cannot be used for data storage of users,   some part of it may be used for some other purpose, e.g.,  software development. This means that all the cloud computing systems go on by an excess of storage. Some scaling parameters are also used   to calculate the   storage/demand available in the owner/user section. These are namely, $\alpha$ for the  owner and $\xi_i$ for the $i$-th user.  
\item \textit{Initial storage: } There are some initial storage for distribution of data in the owner section, i.e.,    $\alpha v_c^{(0)}$ and   in the  user section, i.e., $\xi_ix_i^{(0)}$ for the $i$-th user.
\end{itemize}
 Using the model  \eqref{eq-model},    one can calculate  how much data is required   by the user's section  and  how much data  can be distributed by the owner section to the users in a particular interval of time (\textit{cf}. Table \ref{table}).   
 \begin{table*} {\scriptsize
\begin{tabular}{|p{21pt}|p{37pt}|p{26pt}|p{35pt}|p{36pt}|p{30pt}|p{35pt}|p{22pt}|p{35pt}|}
\hline
\multicolumn{3}{|c|}{\parbox{77pt}{\centering 
{Parameters}
}} & \parbox{35pt}{\raggedright 
{ Initial storage $\left(\alpha v_c^{(0)}\right)$  of owner}
} & \parbox{35pt}{\raggedright 
{Initial storage  $(\xi_ix_i^{(0)})$  of an user  for $i=1,2$}
} & \multicolumn{2}{|l|}{\parbox{60pt}{\raggedright 
{Storage distribution at $l$-th stage by the owner $\left(\alpha v_c^{(l)}\right)$}
}} & \multicolumn{2}{|l|}{\parbox{60pt}{{
Storage usage at $l$-th stage by an user    $\left(\xi_ix_i^{(l)}\right)$}
}} 
\\  \cline{1-3}  $\alpha$   &  $\xi_1$   & $\xi_2$   &  & \\
\cline{1-3} \cline{4-9}

\parbox{21pt}{\raggedright \multirow{5}{*}{
{$0.6$}}} & \parbox{21pt}{\raggedright \multirow{5}{*}{
{$1.25$} }} & \parbox{21pt}{\raggedright \multirow{5}{*}{
{$1.28$} }}
 & \parbox{35pt}{\raggedright \multirow{5}{*}{
{$1$ Gb}
}} & \parbox{36pt}{\raggedright \multirow{5}{*}{
{$0.1$ Gb}
}} & \parbox{21pt}{\raggedright \multirow{2}{*}{
{$l=1$}
}} & \parbox{30pt}{\raggedright \multirow{2}{*}{
{$480$ Mb}
}} & \parbox{13pt}{\raggedright 
{$i=1$}
} & \parbox{31pt}{\raggedright {$12.29$ Mb}
}
\\
\cline{8-9} 
 &    &  &  &  &  &  & \parbox{13pt}{\raggedright 
{$i=2$}
} & \parbox{31pt}{\raggedright {$13$ Mb}
} 
\\
\cline{6-9} \cline{8-9} 
 &    &  &  &  & \parbox{21pt}{\raggedright \multirow{2}{*}{
{$l =10$}
}} & \parbox{22pt}{\raggedright \multirow{2}{*}{
{$369$ Mb}
}} & \parbox{13pt}{\raggedright 
{$i=1$}
} & \parbox{31pt}{\raggedright 
{$103$ Mb}
} 
\\
\cline{8-9} 
 &  &    &  &  &  &  & \parbox{13pt}{\raggedright 
{$i=2$}
} & \parbox{31pt}{\raggedright 
{$102.76$ Mb}
} 
\\
\cline{6-9} 
 &      &  &  &  & \parbox{21pt}{\raggedright \multirow{2}{*}{
{$l=20$}
}} & \parbox{22pt}{\raggedright \multirow{2}{*}{
{$3.5$ Gb}
}} & \parbox{13pt}{\raggedright 
{$i=1$}
} & \parbox{31pt}{\raggedright 
{$1.16$ Gb}
} 
\\
 \cline{8-9}
 & &  
 &      &  &  &  & \parbox{13pt}{\raggedright 
{$i=2$}
} & \parbox{31pt}{\raggedright 
{$1.169$ Gb}
} 
 \\
\cline{6-9} 
 &      &  &  &  & \parbox{21pt}{\raggedright \multirow{2}{*}{
{$l=200$}
}} & \parbox{25pt}{\raggedright \multirow{2}{*}{
{$10.45$ Gb}
}} & \parbox{13pt}{\raggedright 
{$i=1$}
} & \parbox{31pt}{\raggedright 
{$1.73$ Gb}
} 
\\
\cline{8-9} 
 &      &  &  &  &  &  & \parbox{13pt}{\raggedright 
{$i=2$}
} & \parbox{31pt}{\raggedright 
{$5.03$ Gb}
} 
 \\
\cline{6-9} 
 &      &  &  &  & \parbox{21pt}{\raggedright \multirow{2}{*}{
{$l=365$}
}} & \parbox{22pt}{\raggedright \multirow{2}{*}{
{$7.07$ Gb}
}} & \parbox{13pt}{\raggedright 
{$i=1$}
} & \parbox{31pt}{\raggedright 
{$4.02$ Gb}
} 
\\
\cline{8-9} 
 &   &  &  &  &  &  & \parbox{13pt}{\raggedright 
{$i=2$}
} & \parbox{31pt}{\raggedright 
{$123.5$ Mb}
} 
\\
\hline
\end{tabular}
\caption{An example of a cloud computing system with one owner and two users. The symbols $\alpha$, $(\xi_1,~\xi_2)$ denote the scaling parameters corresponding to the capacity and demand variables $v^{(l)}_c$ and   $x^{(l)}_i$ respectively. }
\label{table}}
 \end{table*}
\par
The cloud computing model \eqref{eq-model} is such that it exhibits chaos in the distribution of storage between the owner and users. So, the data storage (in the cloud)   to the owner section  may be random   which ensures that the database system   allocates  data randomly as per  users' demands.  Thus, the proposed model is capable of preventing external threats like   data stealing and   it secures data privacy. Next,  we discuss about some other issues, namely   user authentication, data security and other external threats.
\\ \textit{1. User authentication:} One of the most security concerns in cloud computing is the user's authentication   to the owner.   Here, the process of verifying the user's identity to the owner consists of two steps: 
\begin{itemize}
\item \textit{User identification: } This step presents an identification for each and every user, and  the user section carefully preserves its identity (ID) and password which are  provided by the owner. These ID and password must be kept secretly and  known to the user section only. Otherwise,  the  data can be accessed by some other one or data can be stolen. 
\item \textit{Verification by the owner section:}  A verification of ID and password is made by the owner which is necessary to allocate data to each user distinctively. To verify the ID and password, owner section uses secured hash algorithm (SHA). When an user logs in into the system  with its own ID and password, the owner section  applies the SHA and matches with the user's ID and Password. If it matches then user can access, otherwise it will be denied by the owner section. Also,    each user  is   connected  with the owner section using a unique media access control (MAC) address as well.
\end{itemize}
  \textit{2. Data Privacy and security}:  Data privacy, also called the information privacy, is an important  aspect of information technology (IT) that deals with the ability of an organization or individual   to determine what data in a computing system can be shared with the third  parties.  The main issue is the trust on the user and also on the owner.  So, in this manner owner should develop the user's data privacy. The users and  owner  all are connected through a network, and users  require to create nodes to save data as per the storage distribution by the owner. So, in  each node the data must be securely encrypted and stored. In this way, one can use AES-$256$  encryption scheme \cite{stallings2015} for data encryption and decryption \cite{roy2017,roy2019} by which the user can know the private key to access their files when it is needed. Also as the process of data transfer occurs in the cloud with superior encryption procedure, the data is not corrupted or tampered in any way.
\par  On the other hand, keeping all cryptographic algorithms up to date is an another factor when the information needs to be protected  by encryption. In this case, users must use   advanced cryptographic algorithms that are up-to-date which deny faulty algorithms of old cryptographic process.  This certainly prevents   the information leakage for the owner  and users. If the owner is not aware of or concerned about these changes, the security risk will continue to increase, while attackers are looking for specific vulnerabilities dealing with unsuitable cryptographic algorithms.\\
 \textit{3. External threats}:   An owner and users must be aware of the  external threats that an attacker  exploits the vulnerabilities in services provided to an user. 
  External threats can be characterized by attacks that occur outside the user's domain.  External threats include brute force attack, hacking of data, hacking of user ID and password  etc.  The attacks occur    when a hostile user deploys a proxy application in between an user and owner without them knowing, and the attacker intercepts personal information such as user ID and password.   
\par      In order to avoid brute force attack in our proposed model we use AES-$256$  encryption scheme. The encryption occurs in two ways--one by the owner section in which the data is stored in the cloud and other by the user section where the data is stored at the physical level. In Sec. \ref{sec-sub-data-loss} we have analyzed the  process  of data loss and how it can be minimized in the cloud computing system. For this purpose  we construct nodes for data storage where each device  in the user section contains $50$\% of the total data that are stored in nodes  in such a way that a single block of nodes does not contain  full data information, implying that  the data is safe, i.e.,     an attacker can not recover any information without collection of full data.  Again in Sec. \ref{sec-traffic-load} for the analysis of traffic load,  we discuss about chunks where the data storage is  divided into different parts. So, to recover a   data from nodes while the data is   transferred, an attacker has to recover all the chunks. However,  in the process of creating chunks of files, users always use the initial condition for the capacity variable and it is almost impossible for an attacker to choose the initial condition. Furthermore, since the data is in transit  by parts then it is also impossible to get all the parts of the storage nodes.    Thus,  this approach, not only safeguards data where it is stored, but also helps assure users that data is secured while in transit. \\
 \textit{4. Unauthorized server}:   As the data needs to be transmitted over a network to the owner's cloud, there are numerous means through which an attacker can easily get into the Internet based network and act as an user to the owner's  data, thus resulting into the loss of data. To prevent the loss of data, a well developed authorized certification may be used. Here, Certificate Authorities (CAs) issue each certificate, which is a credential for the online world, to only one specific domain or server. The cloud server first sends the identification information to the owner when it connects, then sends the owner a copy of its  certificate. The owner verifies the certificate and then sends a message to the server and the server sends back a digitally signed acknowledgement to start an  encrypted session, enabling encrypted data transfer between the owner and the user. Moreover, the data and keywords are stored on the cloud in encrypted form.
\subsection{Analysis of data loss} \label{sec-sub-data-loss}
Replication scheme has been widely used as a means to achieve high availability and to avoid failure of data in distributed cloud storage systems \cite{qu2012,chun,koo2011,chansler2012,dean2010}. However, in most of the existing schemes, e.g., random replication scheme, copyset replication scheme, HDFS random placement policy, the probability of data loss  may not be reduced. Here, we propose a different replication scheme which is resilient, fault-tolerant and high-efficient global replication algorithm (RFH) for distributed cloud storage systems. Before proposing a scheme for the calculation of the probability of data loss, we note that  in order to maintain a low-cost of data storage for users, owners may choose a pseudo-randomness system of partitioning their storage as per the  demand of users. These partitions are random in storage and may be called  as nodes, i.e., the $l$-th node for the $i$-th user  may be defined as  $N_i^{(l)}= \xi_ix_i^{(l)}$.  
\par    
It is to be mentioned that a cloud storage system may have  some limitations in physical   (IaaS) as well as software  (SaaS) levels, and so it may not provide an effective performance in between the owner and users   without a well-designed storage allocation policy. In fact, there is always an unpredictable demand of storage in the owner's section by the users owing to customer services.  As the  demand rate of an user in any  interval of time or stage becomes higher and hence the storage in the  owner's section,  the  chance of machine failure or device crush  increases. Thus, proper design of a replication scheme is most important to save the  storage nodes or   data of users. 
\par
The traffic load or cluster-wide power outage is one of the main concerns for data loss in a cloud storage system unless any replication of nodes for data storage is done.  The replication of nodes can restore the data again and repair the affected nodes. Sometimes the power outage can kill a data in storage nodes. Even after when the power is restored, the nodes can not be recovered. However,  in our proposed scheme,  the replication of nodes gives much lower probability of losing data from the affected nodes than the existing schemes \cite{cidon2013}. Before providing a replication scheme we review some main issues of data loss and how our model is useful for avoiding data loss as follows:    

 \subsubsection{Traffic load in cloud computing system} \label{sec-traffic-load}   Network traffic  control is   used to control the load of data traffic in which the data is transferred from users to owner and vice versa in a given interval of time or at any stage (say, $l$-th iteration) of the computing system.     In our proposed model \eqref{eq-model}, since the allocations  of data for every users are given as per their demand rates, the possibility of traffic load is low. However, there are some other traffic issues, namely the consumption of data bandwidth during working hours, bandwidth competition to purchase spectrum etc. which can affect the networking system.  This issue can be resolved   by   creating chunk files in the storage nodes of data traffic due to the demands  of users. If consumption of data bandwidth occurs in the user section in any node, then it is required to partitioning  the allocated storage in the respective nodes  which  reduces  the traffic problem. Below we show how the chunk files in the nodes may be created.
  \par
 The demand rate  of an $i$-th user at the $l$-th stage is given by  $q_i^{(l)}= \xi_i x_i^{(l)} v_{c}^{(l)}$, where $l=1,2,...$.  Since the forward demand rate creates some traffic,   the demand of an user at $l$-th stage can be written as the difference of the demands during the traffic at $l'$-th stage and that at the $l''$-th stage, i.e., 
   $x_i^{(l)}=x_i^{(l')}-x_i^{(l'')}$ with     $l'\leq l \leq l''$, so that the storage  in  the  $l$-th node of an $i$-th user is  $\xi_i x_i^{(l)}$.
 Next, to calculate the number of chunk files to be created in the node $N_i^{(l)}$  in the  interval $ l'\leq l \leq l''$, where  each $l$ may not be equally spaced,  we note that the size of the storage in the $l$-th node of an $i$-th user  is $c_i^{(l)}$= $\xi_ix_i^{(l)}$ where  $l'\leq l\leq l''$. So,  at $l=0$,   $c_i^{(0)}$ represents the  initial chunk size for a node. The chunks in the next stages are  calculated as $c_i^{(l+1)}=c_i^{(l)}-c_i^{(l)}v_c^{(l)}$, so that   $c_i ^{(l)}$ represents the total number of chunk files in the node $N_i^{(l)}$ for an $i$-th user.  
 \subsubsection{Cluster-wide power outage} \label{sec-power-outage}  The data loss in cloud computing  may be due to the cluster-wide power outage. Here, a cluster means a collection of interconnected   different machines (devices or computers) which store  all the chunk files that  contain  the total users'
 data in a node.  The cluster-wide power outage causes the failures of the machines in the cluster or crush of the devices where the database is being stored.  In  a cloud computing system there is a connection between the physical storage in a device and the virtual storage in the cloud. In our model the user section has nodes in physical level and uploads its data to the owner section's   cloud storage. If in a cluster all the   machines    fail, there is a possibility of   nodes failure at physical level, however, the data may be recovered from the cloud storage.  Note that the data may also be  lost or crushed in the cloud   due to some other reasons, e.g., virus attack.    To resolve these issues we propose a design for preventing data loss as follows:
\par Having  calculated nodes for each user  and the chunk files for each node, we redistribute the chunk files  into  one primary and   two secondary nodes in such a way that in the first replica $100\%$ of the chunk files remain in the primary node and in the second replica $50\%$ of the chunk files are in one secondary node and rest $50\%$ in the other.   Here,  each node   is   stored in the owner section (cloud storage) as well as in the users' section (physical device).    
 Let us now consider $n$ number of nodes, say $\{1,2,3,...,n\}$. The distribution of nodes is given in Table \ref{table1}.
\begin{table*}
\begin{tabular}{|c|c|c|c|c|} \hline
{Primary (P)}&{ Secondary-1~($S^1$)}& {Secondary-2~($S^2$)}&{ Nodes in the owner section}&{ Nodes in the user section}\\ \hline
{$P_1, P_2,$}& {$S^1_1,S^1_2,$}&{$S^2_1,S^2_2,$}&{$ \{P_1,S^1_2,S^2_3\}$,$ \{P_2,S^1_3,S^2_4\}$,}&{$ \{S^1_1,S^2_2,S^1_3\}$,$ \{S^1_2,S^2_3,S^1_4\}$,}\\ 
 {$...,P_n$}& {$...,S^1_n$}&{$...,S^2_n$}&{$...,\{P_n,S^1_1,S^2_2\}$}&{$...,\{S^1_n,S^2_1,S^1_2\}$} \\ \hline
\end{tabular}
\caption{Replication scheme is shown for user and owner sections with primary $(P's)$ and secondary $(S's)$ nodes. }
\label{table1} 
\end{table*}
The distribution of nodes in the  owner section is  such   that 
  $N_i=\{P_iS^1_{i+1}S^2_{i+2} \}$ for $i=1,2,3...n$  and $S^1_{n+1}=S^1_{1},~S^2_{n+2}=S^1_{2}$, i.e.,   the process is cyclic to fill the node blocks  properly to prevent data loss. Here, $P_i$ and $S_i$ stand for the primary and secondary nodes.  Similarly,   in the user section   to  fill the node blocks we take the secondary data values in such a way that   $N_i=\{S_1^iS_2^{i+1}S_1^{i+2}\}$  and $S^1_{n+1}=S^1_{1},~S^2_{n+2}=S^1_{2}$. Thus,  the number of blocks in the owner's and user's sections is each $n$.  
\subsubsection{\textit{Probability of data loss}}  We note that  the data is   replicated three times one in primary nodes and two in secondary nodes.  If  a machine or device can store    $50\%$ of the total chunk files in the nodes then  for each block in the owner section we require four machines to store the data  and in the user section we require three machines. To calculate the probability of data loss, we assume that the probability of machine failure is $p$.  
\par
 Since the number of blocks is twice the  number of nodes, i.e., $2n$, and two blocks in the user section and owner section have three plus four, i.e., seven machines, there are $7n$ machines for $n$ nodes in two racks (one is for owner and other is for the user).   We calculate the probability of data loss  for $f$ number of  machine failures randomly in both the racks where  
   $n$ represents the number of nodes,  $b~(=2n)$ the  number of total blocks,    $7n$ the number of machines in a cluster,  and $p$ the probability of machine failure.
 \par
  We mention that the above replication scheme is valid for $n\geq 3$ and in order to have data loss there must be failure of  at least three machines at a time. 
 In what follows,  we construct a generating polynomial from Table \ref{table1} and obtain the coefficients of this polynomial to calculate the probability of  data loss.  The generating polynomial is given by
 \begin{equation}
P(x)=\left(1+a_1x+a_2x^2+a_3x^3+a_4x^4+a_5x^5\right)^n,
\end{equation}
  where $a_1={{7}\choose{6}}=7$, $a_2={{7}\choose{5}}=21$,  $a_3={{7}\choose{4}}-1=34$,  $a_4={{7}\choose{3}}$-${{4}\choose{3}}-1=30$ and $a_5={{7}\choose{2}}$-${{4}\choose{2}}$-${{3}\choose{2}}=12$. Here, the coefficient of $x^f$ is the number of ways of choosing $f$ machines out of  $7n$ machines, i.e.,  in ${7n}\choose{f}$ ways with the condition as given in Table \ref{table1}. 
     The probability of no data loss for failure of $f$ machines is given by 
     \begin{equation}
      P_{n}=\left\lbrace \begin{array}{cc}
 \left(\mathrm{coefficient~ of}~x^f~ \mathrm{in}~ P(x)\right)/{{{7n}\choose{f}}} & \mathrm{if}~ f\leq 5n  \\ 0   & \mathrm{otherwise}.
\end{array}\right.   
     \end{equation}
 So, the probability of data loss  for failure of $f$ machines is     $ (1-P_{n})\times P_f$, where $P_f$ denotes the probability of $f$ machine failures, i.e.,   $P_f = {{7n}\choose{f}} p^f (1-p)^{7n-f}$ since there is no data loss for  $f=1,2$. 
Note that the coefficient of $x^f$ in $P(x)$ for $f\leq 5n$  can be obtained by considering all the possible combinations of $k_1,~k_2,~k_3,~k_4,~k_5$ which satisfy $k_1+2k_2+3k_3+4k_4+5k_5 = f$. So, the coefficient of $x^f$ in $P(x)$ is $\sum_{(k_1,k_2,k_3,k_4,k_5)}$ ${{{{n}\choose{k_1k_2...k_5}}}}/{{{n}\choose{f}}} a_1^{k_1} a_2^{k_2} a_3^{k_3} a_4^{k_4} a_5^{k_5}$. Thus, if the probability of a machine failure is $p$ then the probability of data loss is given by
   $P_{l}=\sum_{f=3}^{5n}{{7n}\choose{f}} p^f (1-p)^{7n-f} (1-P_{n})+\sum_{f=5n+1}^{7n}{{7n}\choose{f}}p^f (1-p)^{7n-f}$. 
    \par  Next, we compute the probabilities of data loss for some sample values of $n$ and $p$ as given in Table \ref{table2}. Here, we assume that  different cluster of machines has different number of  nodes  $n$. For example, we consider that  $70$, $140$,  $280$, $560$, $700$, $980$ and $1400$ numbers of machines  have    $10$, $20$, $40$, $80$, $100$, $140$ and $200$ nodes respectively.  We also assume that the probability of machine failure is $p=0.01$. 
  \par
\begin{figure*}
    \centering
    \subfigure[]
    {\includegraphics[width=3in,height=2in]{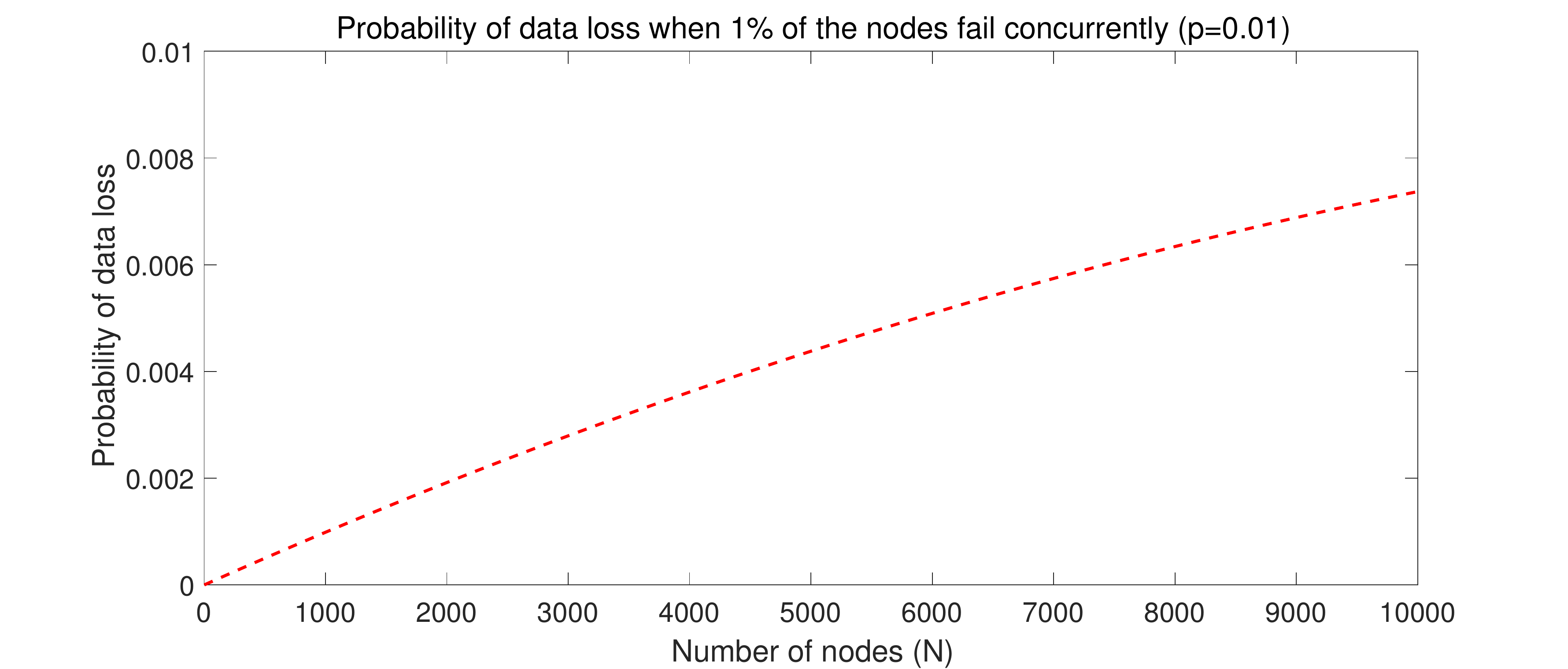}
      }
    \subfigure[]
    {
        \includegraphics[width=3in,height=2in]{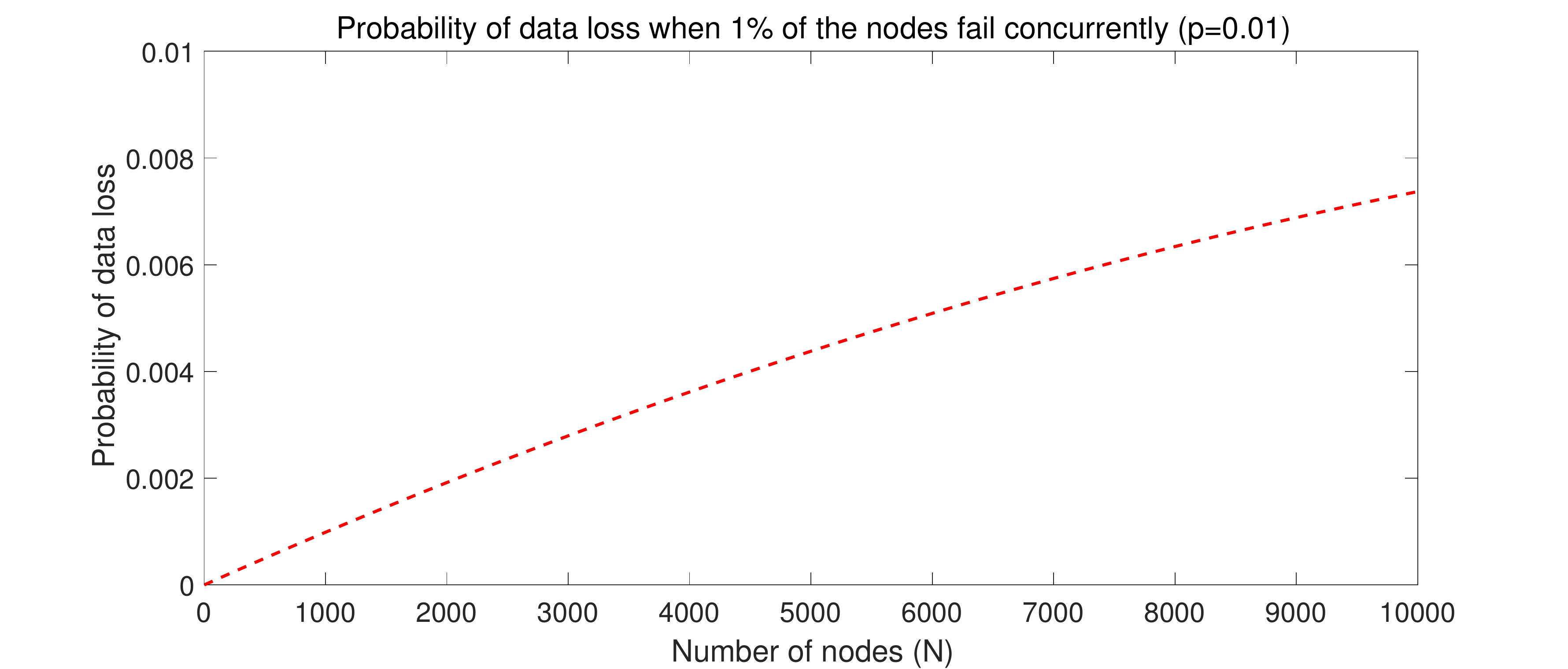}
    }
    \caption{Probability of data loss is shown using our replication scheme [left panel (a)] and compared with that in Ref. \cite{cidon2013} [right panel (b)]. It is seen that   the probability of data loss in our proposed schemes is significantly reduced compared to that in Ref. \cite{cidon2013}.     }
    \label{fig:dataloss}
\end{figure*}

  Figure \ref{fig:dataloss} shows the graphs for the probability of data loss using our proposed scheme [panel (a)] and  as in Ref.  \cite{cidon2013} [panel (b)]. It is found that   in the random machine failures with $p=0.01$, the probability of data loss in our scheme is highly reduced compared to that in Ref. \cite{cidon2013}.
    
    \begin{table*} 
   \begin{tabular}{|c| c|c |c |c}
   \hline
   Number of   & Number of  & Failure  & Probability of \\
   nodes $(n)$ & machines $(7n)$ & policy & data loss $(\times10^{-4})$\\
    \hline
   10	&  70 	& Random	 &	$0.12120$ \\
   
   20     & 140	& Random		&  $0.22220$\\
   
   40	& 280	& Random     & $0.42419$ \\
   
   80	&   560	& Random &$0.82817$  \\
   
   100	&  700  	& Random  &     $1.0301$ \\
   140	&   980	& Random & $ 1.4341$  \\
   200	& 1400	& Random	 & $2.04$\\
   \hline
   \end{tabular}   
   \caption{The probability of data loss is calculated using the random replication scheme as in Sec. \ref{sec-power-outage} for different choices of nodes and machines or devices. The latter are assumed to failure randomly.  }
   \label{table2}           
  \end{table*}       
\section{Conclusion}
We have proposed a discrete dynamical model for cloud computing and management of storage between a third party or company and users. Though the model is applicable for an arbitrary number of users, however, for simplicity, we have analyzed it for one owner and  two users. The basic dynamical properties of the model are studied. It is found that the model   exhibits chaos for  certain ranges of parameter values. A framework for distribution of storage of data and its implementation with users and the owner is also given.   Some issues of data security are analyzed and discussed. Furthermore, we have proposed a  random replication scheme and calculated the probability of data loss. It is found that   the probability of  data loss is highly reduced compared to that using the existing scheme in the literature \cite{cidon2013}.  
  \section*{Acknowledgment}
  The authors thank Anirban Kundu of Department of Mathematics, Visva-Bharati, Santiniketan for some useful discussion. A. Roy and A. P. Misra acknowledge support from UGC-SAP (DRS, Phase III) with Sanction order No. F.510/3/DRS-III/2015(SAPI). 

\end{document}